\documentclass{icsm}          % for LaTeX 2e
\usepackage{amsfonts,amsmath,amssymb}
\usepackage{graphics}

\begin{document}

\title{    On $sl(N)$ and $sl(M|N)$ integrable open spin chains}
\authori{\underline{D. Arnaudon}, 
N.~Cramp\'e, A.~Doikou, L.~Frappat, \'E.~Ragoucy}      
\addressi{Laboratoire d'Annecy-le-Vieux de Physique Th{\'e}orique LAPTH
CNRS, UMR 5108, associ{\'e}e {\`a} l'Universit{\'e} de Savoie
LAPP, BP 110, F-74941 Annecy-le-Vieux Cedex, France}
\authorii{J. Avan}   \addressii{
Laboratoire de Physique Th{\'e}orique et Mod{\'e}lisation
Universit{\'e} de Cergy, 5 mail Gay-Lussac, Neuville-sur-Oise
F-95031 Cergy-Pontoise Cedex
}
\authoriii{}  \addressiii{}
\authoriv{}      \addressiv{}
\authorv{}      \addressv{}
\authorvi{}     \addressvi{}
\headauthor{D. Arnaudon et al.}   
\headtitle{On  $sl(N)$ and $sl(M|N)$ integrable open spin chains}
\lastevenhead{D. Arnaudon et al.:  On $sl(N)$ and $sl(M|N)$ integrable
  open spin chains } 

\maketitle

\begin{flushright}
  \textbf{LAPTH-Conf-1055/04}
\end{flushright}

\begin{abstract}
  We study open spin chains based on rational $sl(N)$ and $sl(M|N)$
  $R$-matrices. We classify the solutions of the reflection equations, for
  both the soliton-preserving and soliton-non-preserving cases.
  We then write the Bethe equations for these open spin chains.
\end{abstract}

\newtheorem{definition}{Definition}[section]
\newtheorem{example}[definition]{Example}
\newtheorem{conjecture}[definition]{Conjecture}
\newtheorem{proposition}[definition]{Proposition}
\newtheorem{theorem}[definition]{Theorem}
\newtheorem{corollary}[definition]{Corollary}
\newtheorem{lemma}[definition]{Lemma}

%% \definecolor{dcyan}{rgb}{0,.8,.8}
%% \definecolor{ddcyan}{rgb}{0,.6,.6}
%% \definecolor{dgreen}{rgb}{0,.8,0}
%% \definecolor{lyellow}{cmyk}{0,0,.2,0}
%% \definecolor{rose}{rgb}{1,.6,.6}
%% \definecolor{violet}{rgb}{.9,0,.6}
\newcommand\textcolor[1]{{}}
\newcommand\noir[1]{\textcolor{black}{#1}}
\newcommand\rouge[1]{\textcolor{red}{#1}}
\newcommand\bleu[1]{\textcolor{blue}{#1}}
\newcommand\green[1]{\textcolor{green}{#1}}
\newcommand\dcyan[1]{\textcolor{dcyan}{#1}}
\newcommand\ddcyan[1]{\textcolor{ddcyan}{#1}}
\newcommand\jaune[1]{\textcolor{yellow}{#1}}
\newcommand\rose[1]{\textcolor{rose}{#1}}
\newcommand\magenta[1]{\textcolor{magenta}{#1}}
\newcommand\violet[1]{\textcolor{violet}{#1}}
\newcommand\dgreen[1]{\textcolor{dgreen}{#1}}

\newcommand{\II}{{\mathbb I}}
\def\CC{{\mathbb C}}
\def\NN{{\mathbb N}}
\def\QQ{{\mathbb Q}}
\def\RR{{\mathbb R}}
\def\ZZ{{\mathbb Z}}
\def\cA{{\cal A}}          \def\cB{{\cal B}}          \def\cC{{\cal C}}
\def\cD{{\cal D}}          \def\cE{{\cal E}}          \def\cF{{\cal F}}
\def\cG{{\cal G}}          \def\cH{{\cal H}}          \def\cI{{\cal I}}
\def\cJ{{\cal J}}          \def\cK{{\cal K}}          \def\cL{{\cal L}} 
\def\cM{{\cal M}}          \def\cN{{\cal N}}          \def\cO{{\cal O}}
\def\cP{{\cal P}}          \def\cQ{{\cal Q}}          \def\cR{{\cal R}} 
\def\cS{{\cal S}}          \def\cT{{\cal T}}          \def\cU{{\cal U}}
\def\cV{{\cal V}}          \def\cW{{\cal W}}          \def\cX{{\cal X}}
\def\cY{{\cal Y}}          \def\cZ{{\cal Z}}
\def\bA{{\bar A}}          \def\bB{{\bar B}}          \def\bC{{\bar C}}
\def\bD{{\bar D}}          \def\bE{{\bar E}}          \def\bF{{\bar F}}
\def\bG{{\bar G}}          \def\bH{{\bar H}}          \def\bI{{\bar I}}
\def\bJ{{\bar J}}          \def\bK{{\bar K}}          \def\bL{{\bar L}} 
\def\bM{{\bar M}}          \def\bN{{\bar N}}          \def\bO{{\bar O}}
\def\bP{{\bar P}}          \def\bQ{{\bar Q}}          \def\bR{{\bar R}} 
\def\bS{{\bar S}}          \def\bT{{\bar T}}          \def\bU{{\bar U}}
\def\bV{{\bar V}}          \def\bW{{\bar W}}          \def\bX{{\bar X}}
\def\bY{{\bar Y}}          \def\bZ{{\bar Z}}
\newcommand{\eps}{{\varepsilon}}
\newcommand{\un}{\mbox{1\hspace{-1mm}I}}
\newcommand{\id}{\mbox{id}}
%%\sfrac12
%%\newcommand{\sfrac}[2]{{\displaystyle{\frac{#1}{#2}}}}
\def\qmbox#1{\qquad\mbox{#1}\quad}
\def\Ad{\mathop{\rm Ad}\nolimits}
\def\tr{\mathop{\rm Tr}\nolimits}
\def\tK{{\tilde K}}
\def\diag{\mathop{\rm diag}\nolimits}

\def\emme{M}
\def\enne{N}

\section{ 
\rouge{ %% 
  Introduction
}}
We are interested in open quantum spin chains based on rational
$R$-matrices of $sl(N)$ and $sl(M|N)$:
\begin{equation}
  R_{12}(\lambda) = \lambda \II+iP_{12}\;,
\end{equation}
where $P$ is the super-permutation operator
\begin{equation}
  \label{eq:Pdef}
  P = \sum_{i,j=1}^{\emme+\enne} (-1)^{[j]} E_{ij} \otimes E_{ji}
\end{equation}
We will give a classification of the reflection matrices compatible with
the integrability of the open spin chain, 
in the two cases of soliton preserving and soliton
non-preserving boundary conditions. 

In section \ref{sect:closed}, we recall graphically the proof of commutation of
transfer matrices for closed chains. We then recall in section \ref{sect:open}
the commutation for open chains.
In section \ref{sect:SNP}, we define the transfer matrix for open spin
chains with soliton non preserving boundary conditions.
In section \ref{sect:RE} we give the classification of solutions of
the reflection equations, for both soliton preserving and soliton non
preserving cases. 
We finally present the analytical Bethe Ansatz method for these chains
and end with the Bethe equations.
More details and references can be found in \cite{selene}.

\section{ 
\rouge{ %% 
  Closed chain integrability
}}
\label{sect:closed}

Let $R$ be a solution of the Yang--Baxter equation 
\bleu{
\begin{equation}
  R_{12}(\lambda_{1}-\lambda_{2})\ R_{13}(\lambda_{1})\ R_{23}(\lambda_{2})
  =R_{23}(\lambda_{2})\ R_{13}(\lambda_{1})\ R_{12}(\lambda_{1}-\lambda_{2})
\end{equation}
}
$$\includegraphics{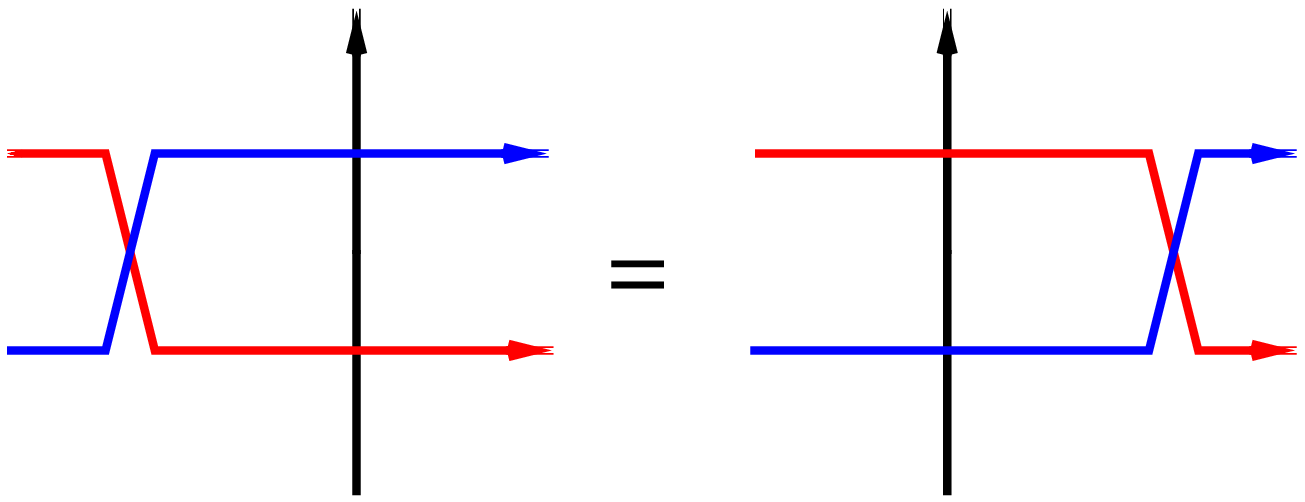}
$$

On a chain with $L$ sites, we define the
monodromy matrix as
\bleu{
\begin{equation}
  T(\lambda) = R_{a1}(\lambda) \ R_{a2}(\lambda) \ \cdots \
  R_{aL}(\lambda) 
\end{equation} 
}
$$\includegraphics{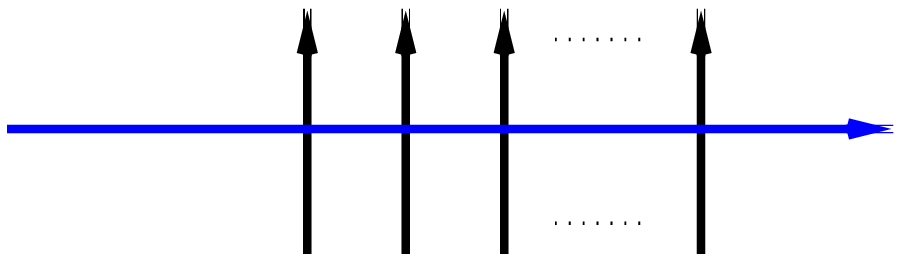}
$$
\\
and the transfer matrix as its \emph{super trace}
\bleu{
\begin{equation}
  t(\lambda) = \tr_a \ T(\lambda)
\end{equation}
}

The Hamiltonian is one of the terms of the expansion of the transfer matrix
\begin{equation}
  {\cal H} = -{1\over 2} {d\over d \lambda}t(\lambda)\Big\vert_{\lambda =0}\;.
\end{equation}

The main property used for integrability
of closed spin chains, i.e. commutation of the
transfer matrix for different values of the spectral parameter,
is the \emph{local} 
Yang--Baxter equation.
\\[3mm]

Graphical proof of the
commutation of transfer matrices $t(u)$ and $t(v)$:
$$\includegraphics{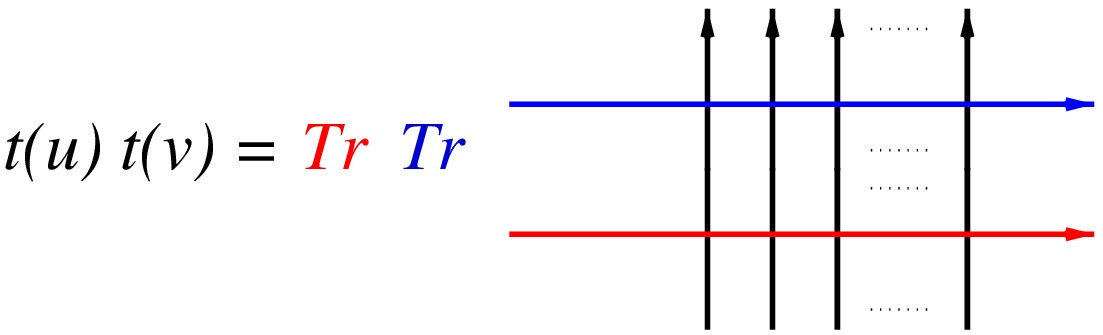}
$$
Insertion of $R_{ab}(u-v) R_{ab}^{-1}(u-v)$ $\longrightarrow$
$$\includegraphics{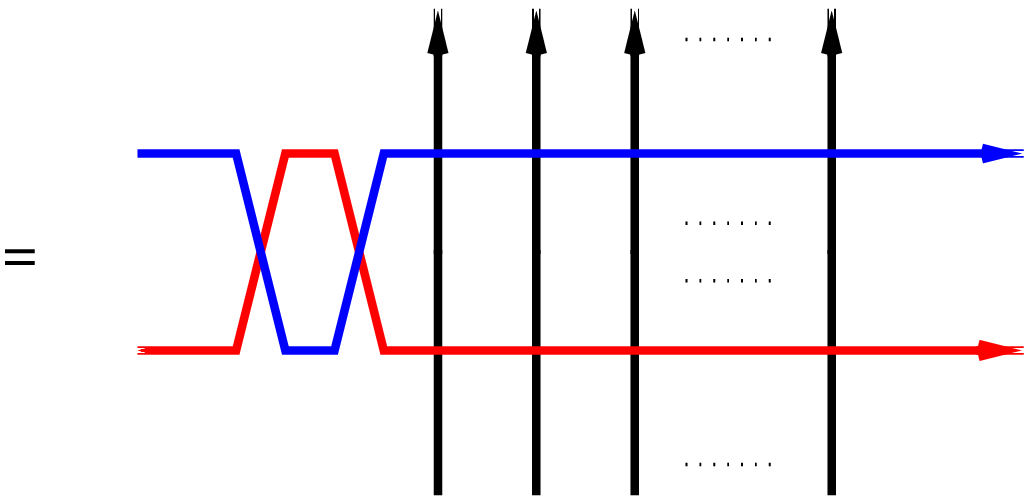}
$$
Use of Yang--Baxter $\longrightarrow$
$$
 \includegraphics{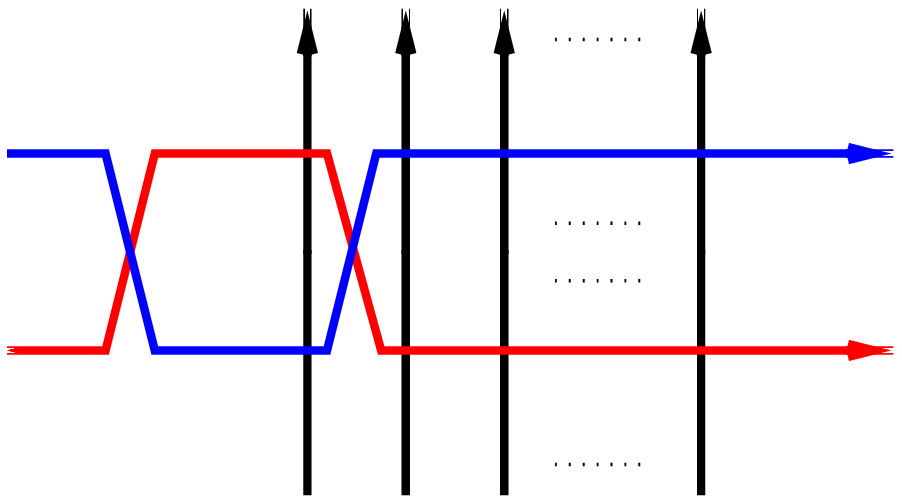}
$$
Use of Yang--Baxter again $\longrightarrow$
$$
\includegraphics{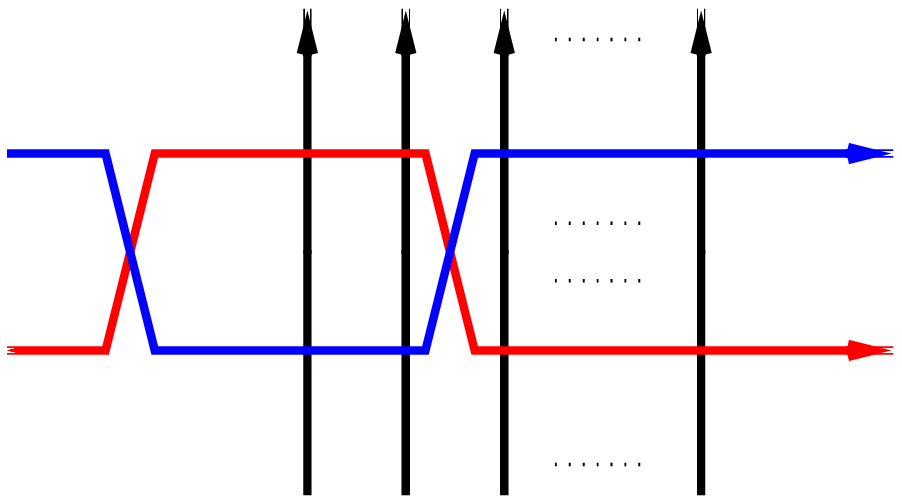}
$$
again...
$$\includegraphics{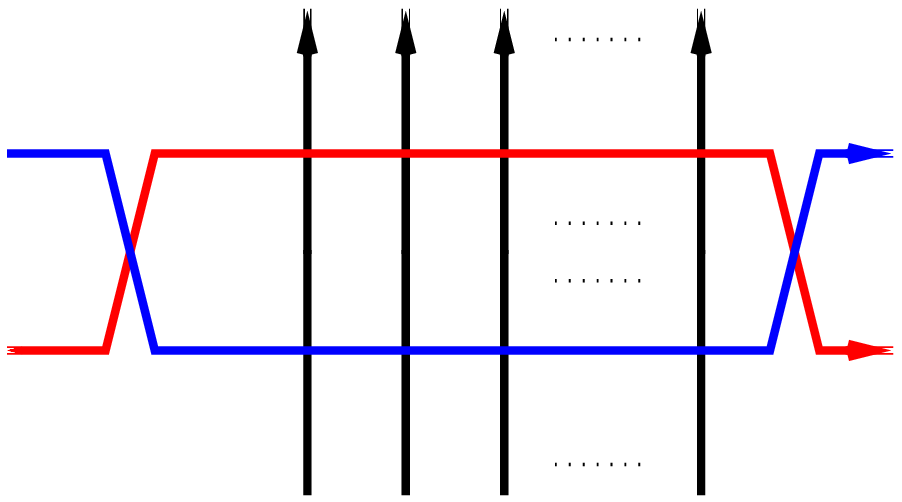}
$$
Cyclicity of trace  $\longrightarrow$
$$\includegraphics{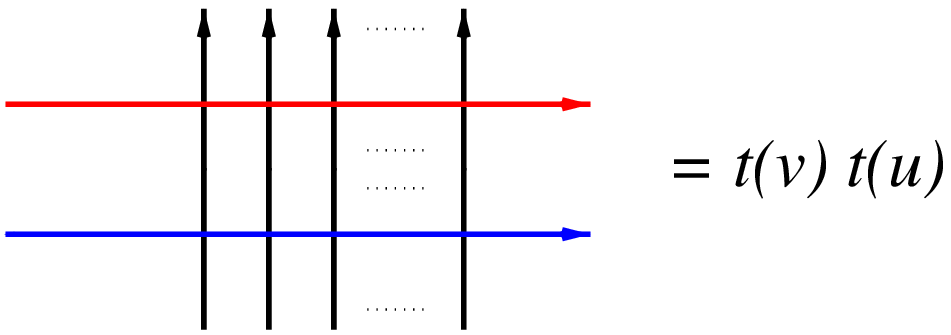}
$$

\section{ 
\rouge{ %% 
  Open chain integrability
}}
\label{sect:open}

For the integrability of open chains, one also needs the 
(local) reflection equation:
\bleu{
\begin{eqnarray}
  R_{ab}(\lambda_{a}-\lambda_{b}) 
  \!\!\!\!&\!\!&\!\!\!\!
  K_{a}(\lambda_{a}) 
  R_{ba}(\lambda_{a}+\lambda_{b})\ K_{b}(\lambda_{b}) \ = 
  \nonumber\\[2mm]
  &&
  K_{b}(\lambda_{b})\
  R_{ab}(\lambda_{a}+\lambda_{b})\ K_{a}(\lambda_{a})\
  R_{ba}(\lambda_{a}-\lambda_{b})
  \label{re}
\end{eqnarray}
}
$$\includegraphics{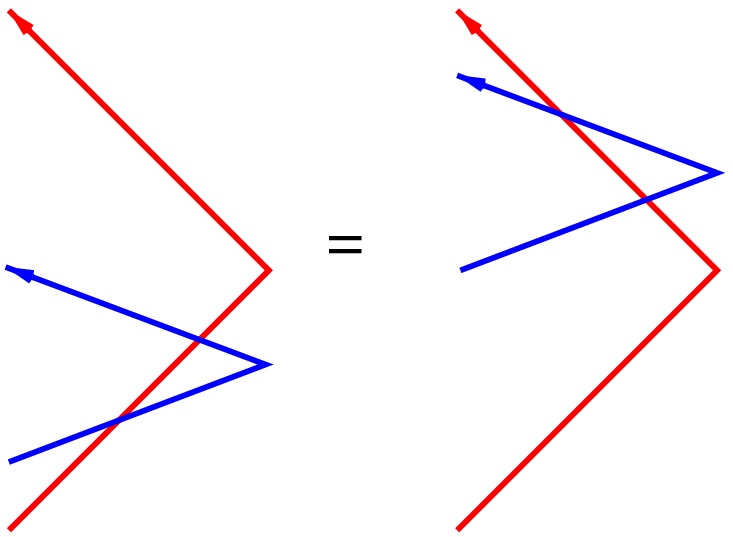}
$$
\\

Let
\bleu{
\begin{eqnarray}
  T_{a}(\lambda) = R_{aL}(\lambda) R_{a,L-1}(\lambda) \cdots R_{a
  2}(\lambda) R_{a1}(\lambda) 
\end{eqnarray}
}
and
\bleu{
\begin{eqnarray}
  \hat T_{a}(\lambda) =
  R_{1a}(\lambda) R_{2a}(\lambda) \cdots R_{L-1,a}(\lambda)
  R_{La}(\lambda) 
\end{eqnarray}
}
The open spin chain transfer matrix is now defined as the super trace:
\bleu{
\begin{equation}
  t(\lambda) = \tr_a K_{a}^{+}(\lambda)\ T_{a}(\lambda)\
  K_{a}^{-}(\lambda)\ \hat T_{a}(\lambda) 
\end{equation} 
}

\bleu{
$t(\lambda)=$
}
$$\includegraphics{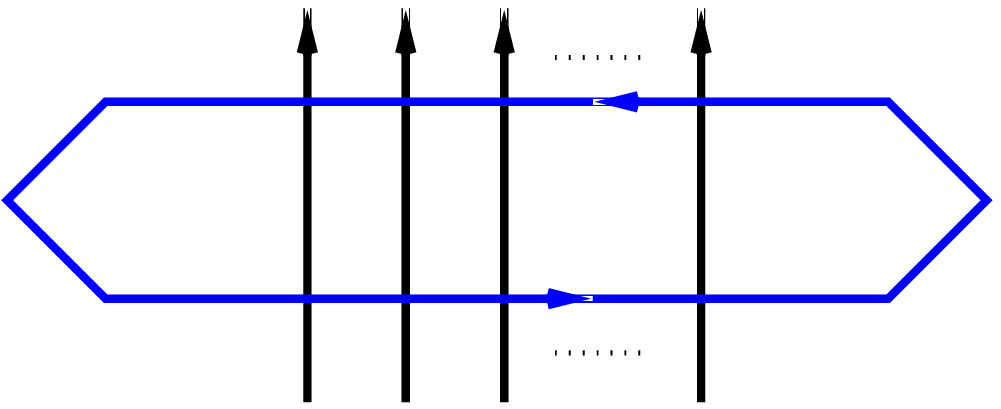}
$$

Graphical proof of the 
commutation of transfer matrices for different spectral \\
parameters, following Cherednik \cite{cherednik} and Sklyanin
\cite{sklyanin}:
\\[1mm]
\bleu{
$t(\lambda_1)\ t(\lambda_2)=$
}

$$\includegraphics{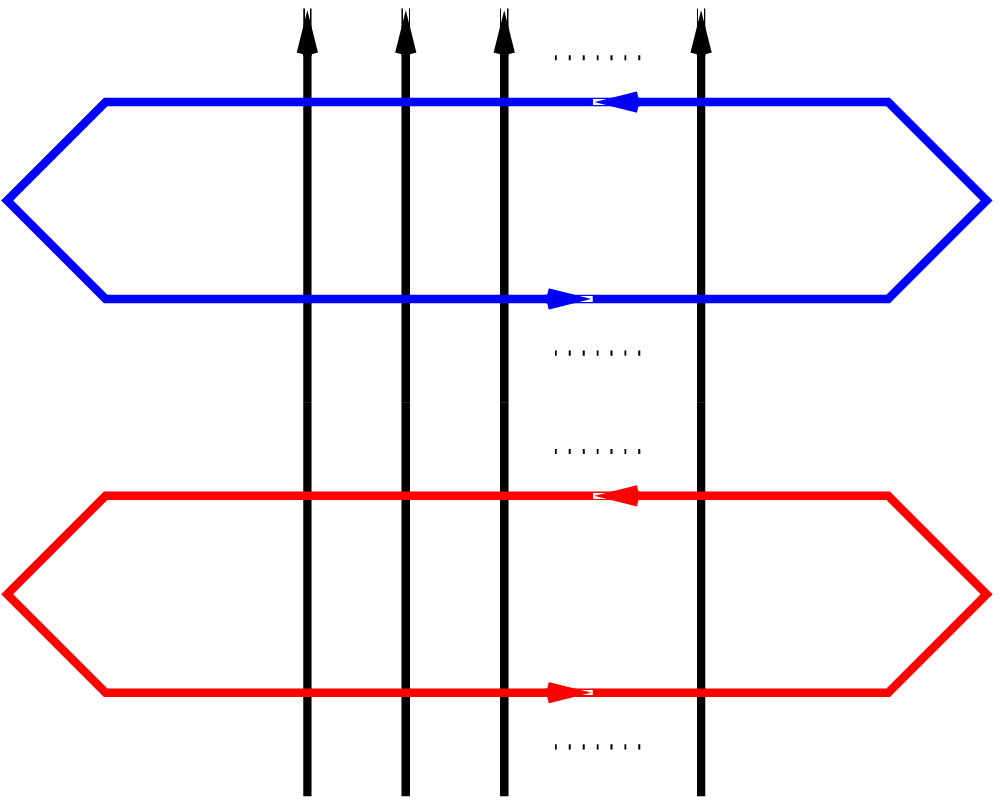}
$$
Insertion of  crossing unitarity :
$$\includegraphics{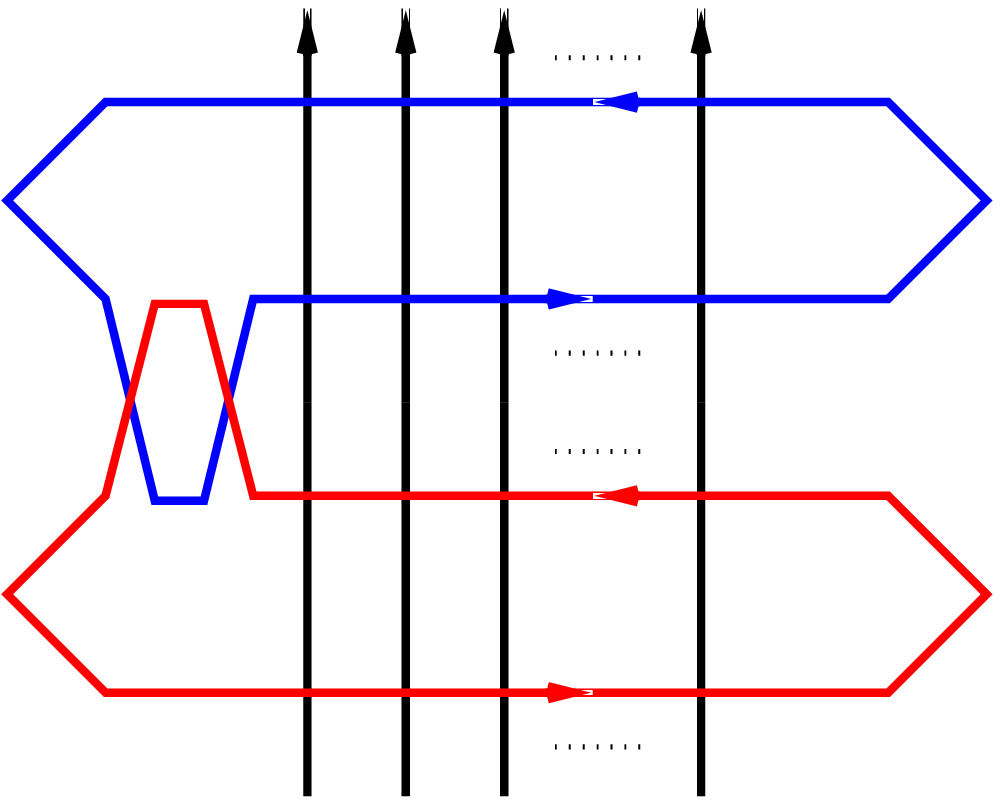}
$$

Use of Yang--Baxter equation :
$$\includegraphics{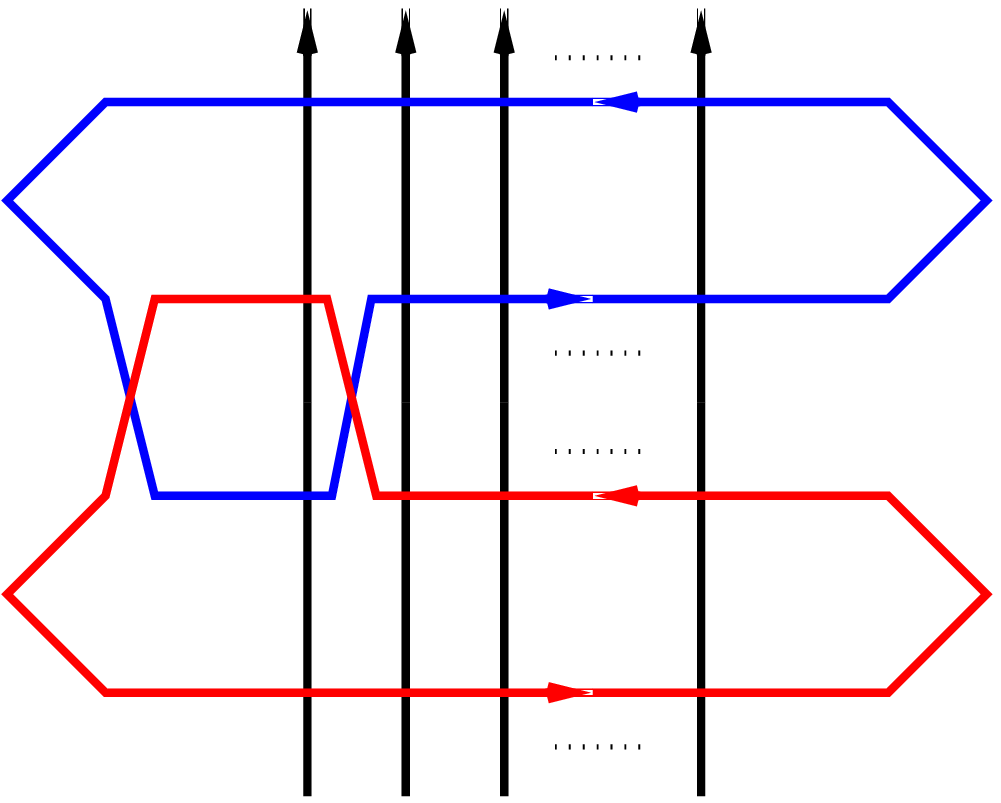}
$$
again and again :
$$\includegraphics{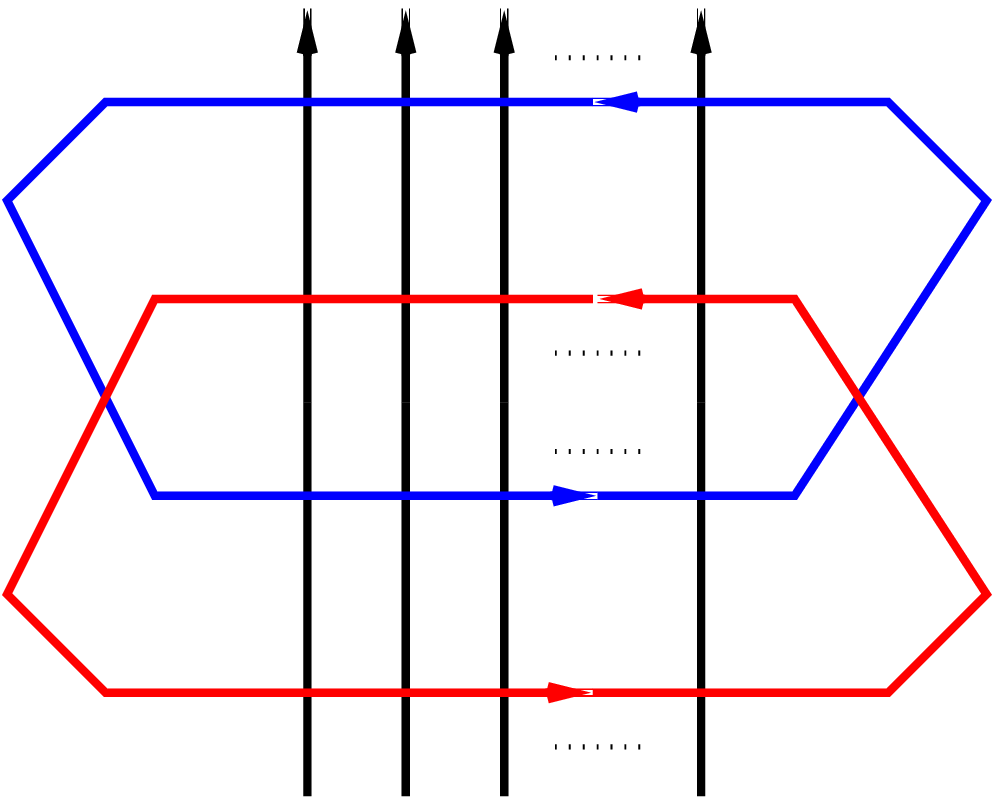}
$$

Insertion of  $R\ R^{-1}$ :
$$\includegraphics{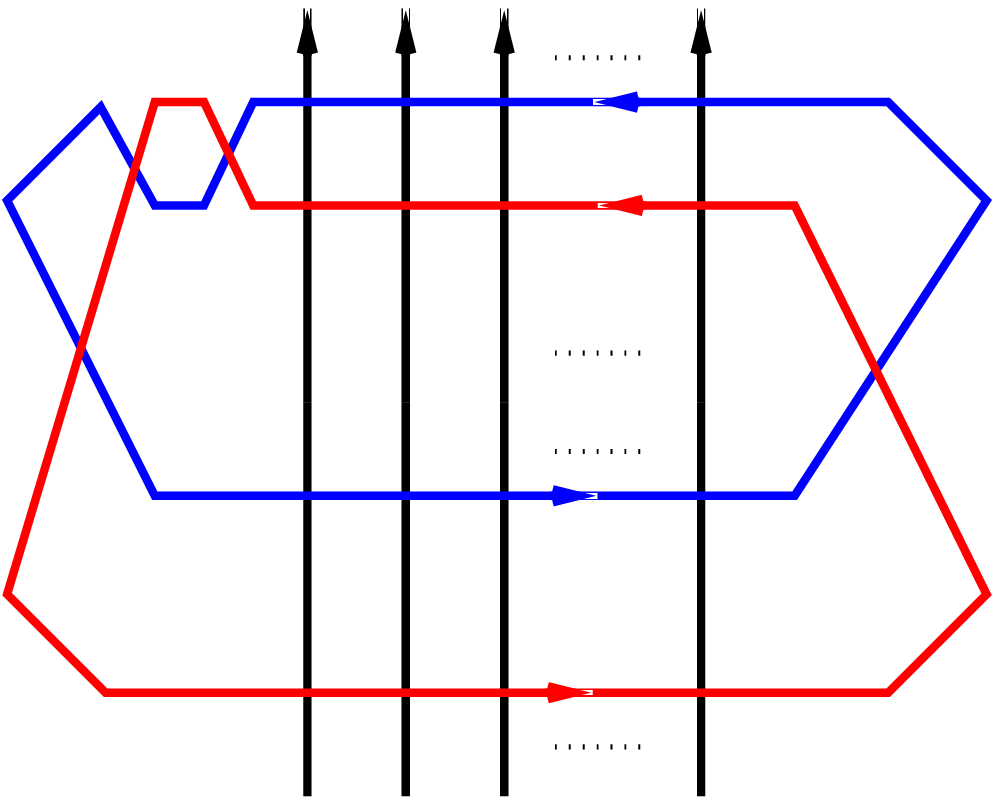}
$$
Use of Yang--Baxter equation :
$$\includegraphics{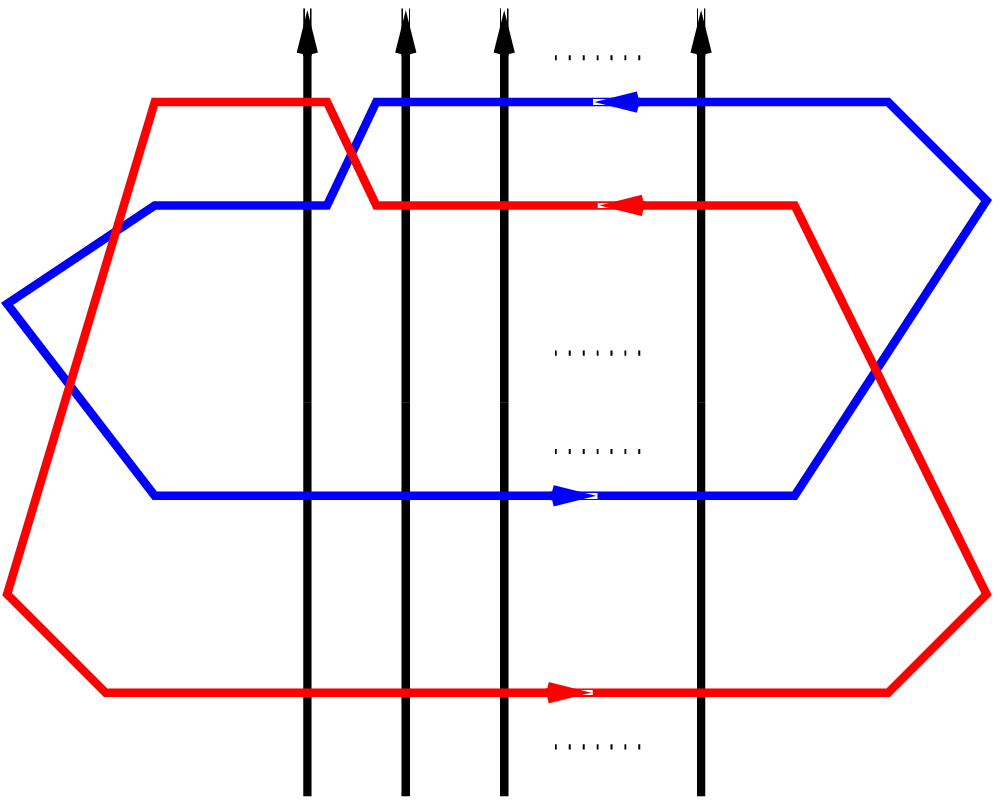}
$$

Use of Yang--Baxter again :
$$\includegraphics{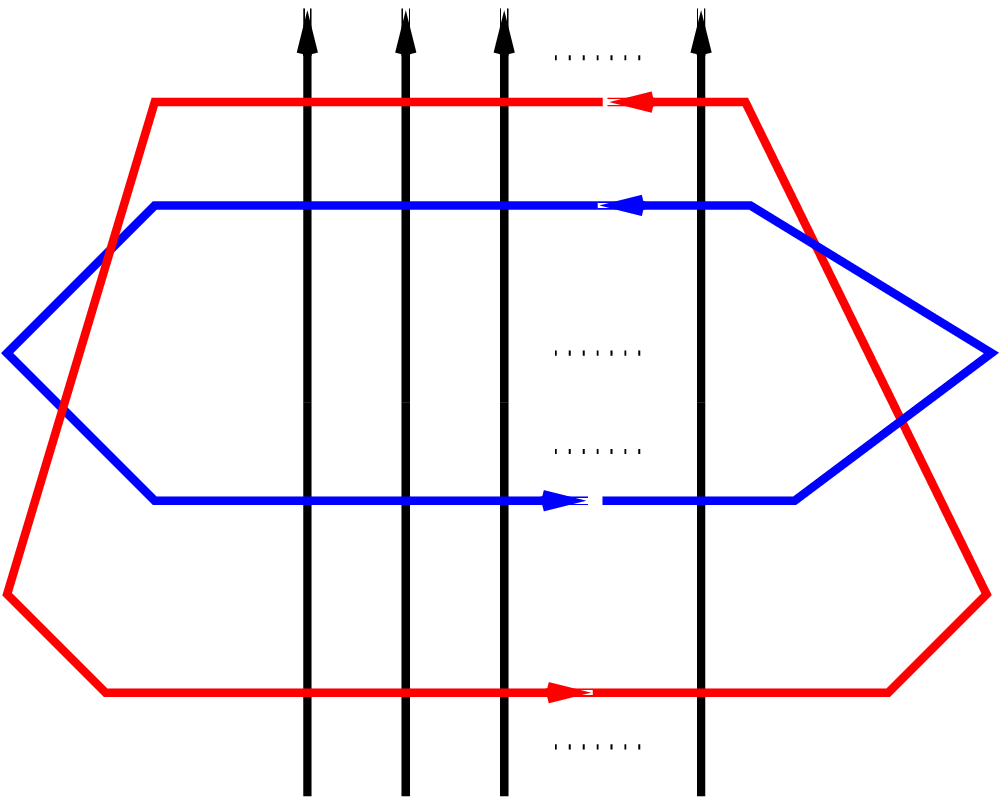}
$$
Use of reflection equation on the left :
$$\includegraphics{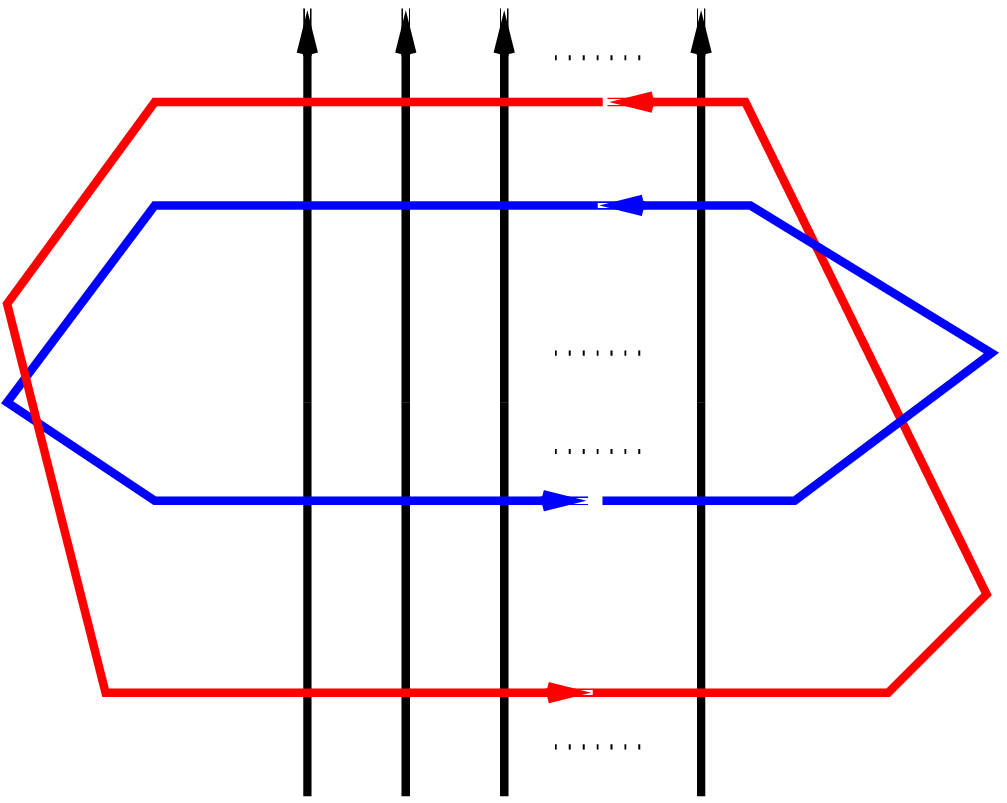}
$$

Use of reflection equation on the right :
$$\includegraphics{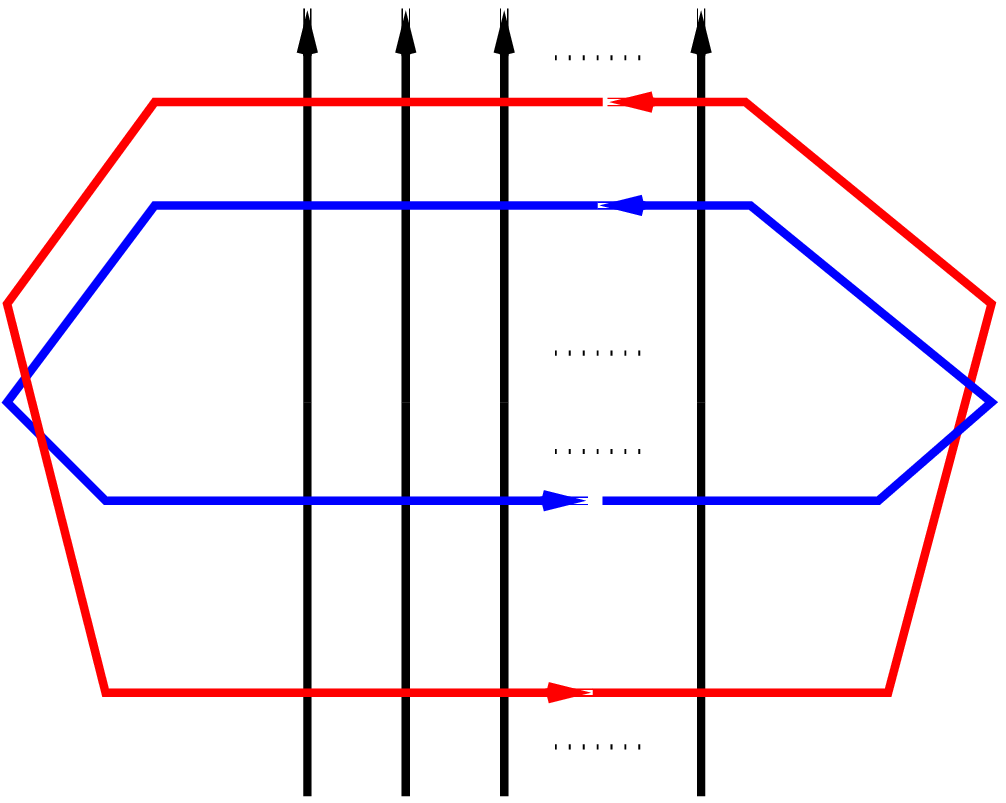}
$$
Use of Yang--Baxter equation :
$$\includegraphics{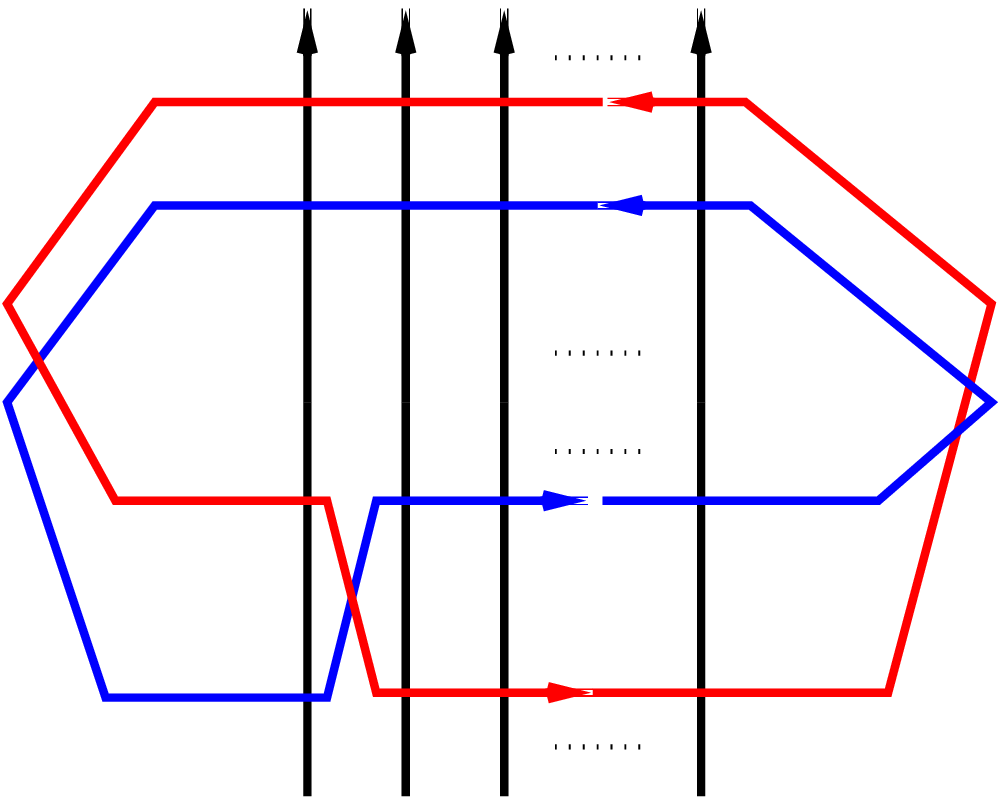}
$$

Use of Yang--Baxter again :
$$\includegraphics{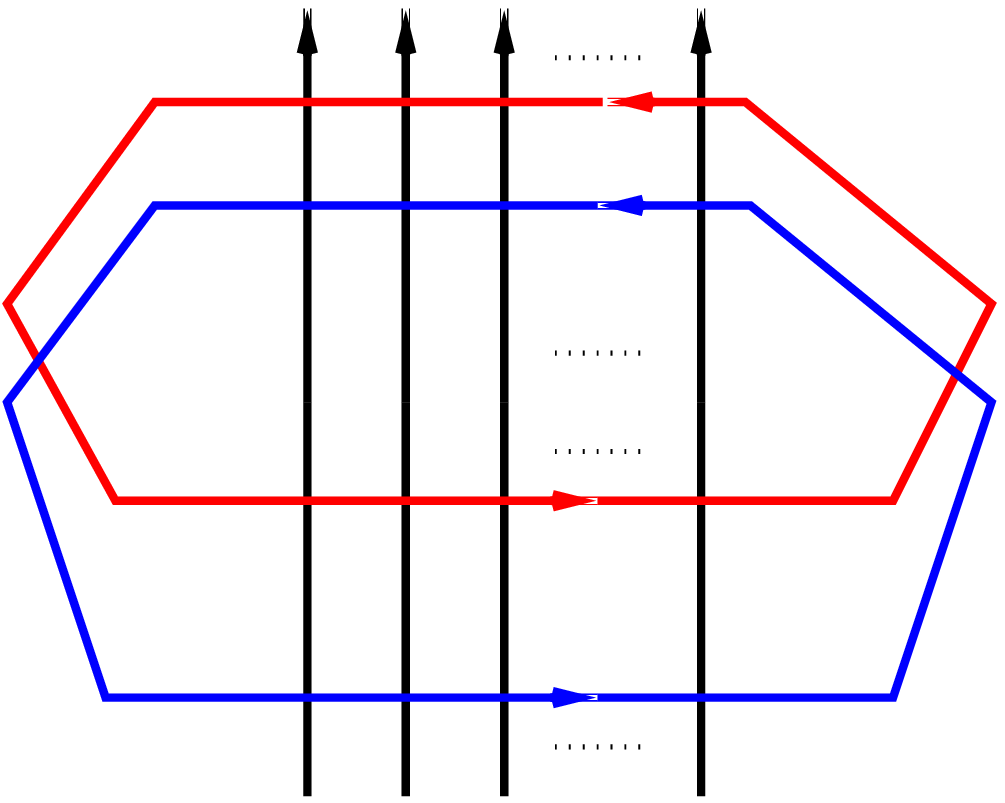}
$$
Another Yang--Baxter :
$$\includegraphics{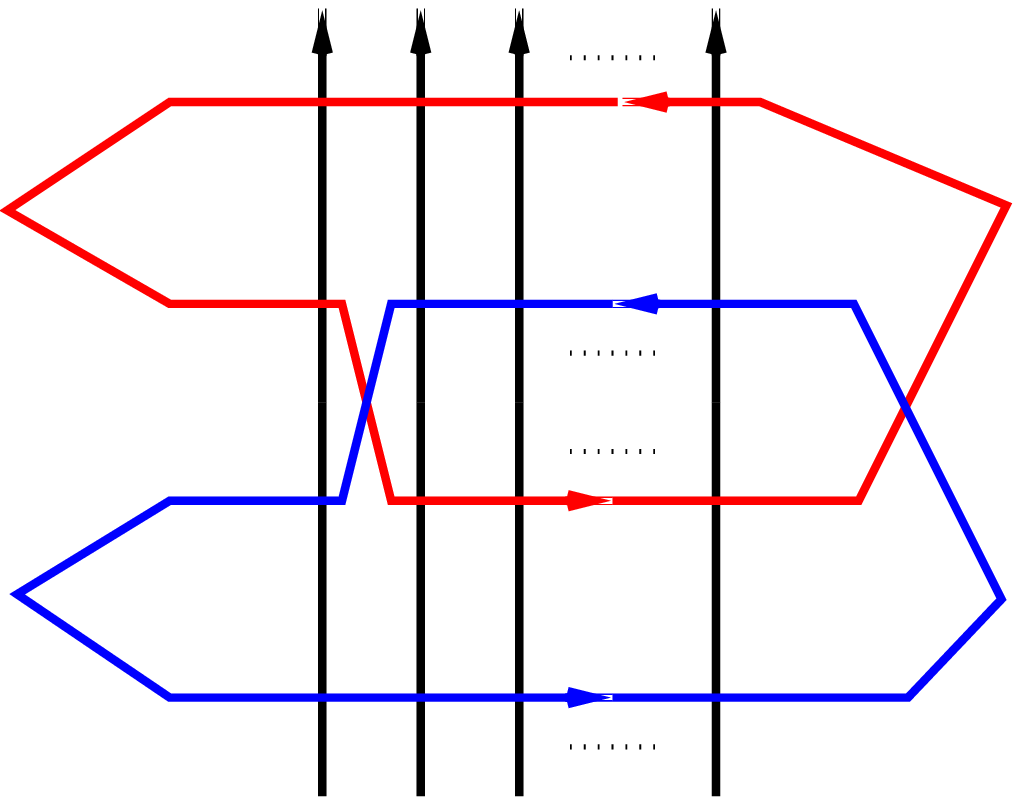}
$$

Yang--Baxter up to the right end :
$$\includegraphics{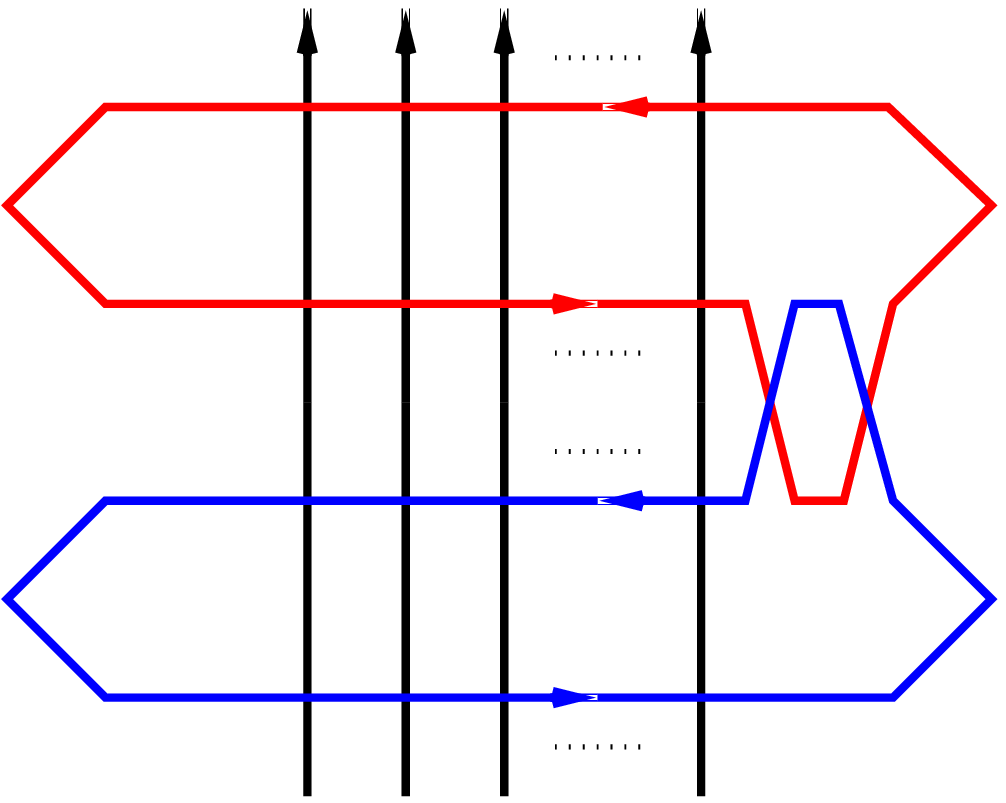}
$$
Use of crossing unitarity :
$$\includegraphics{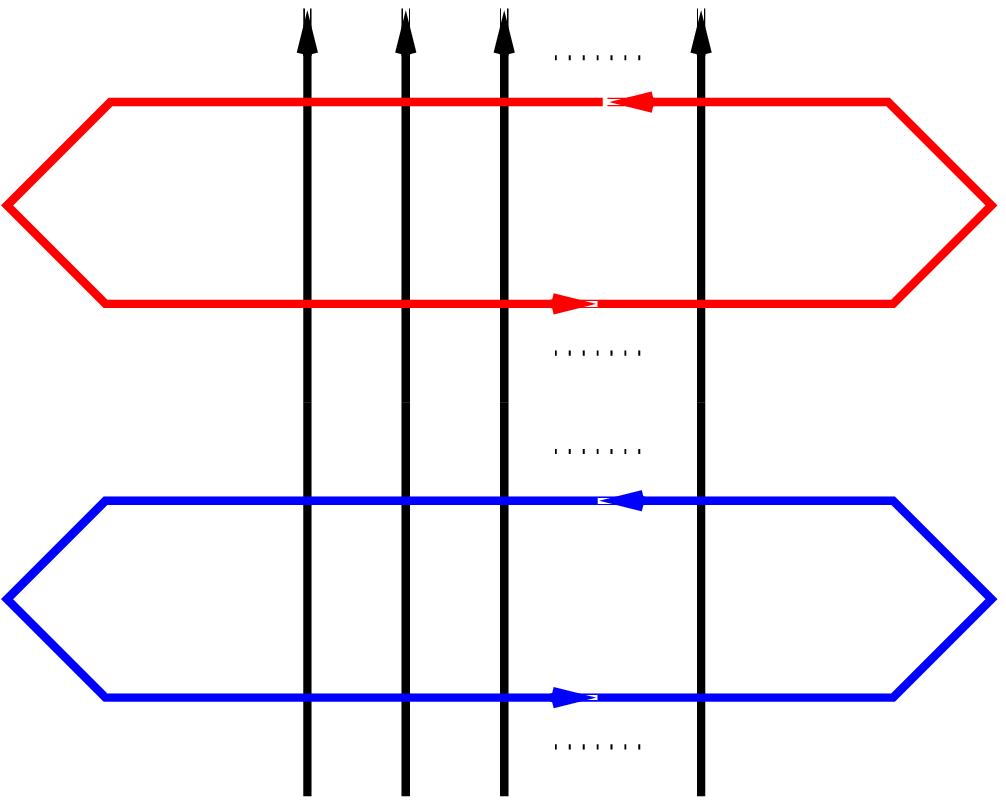}
$$
which is $t(\lambda_2) \ t(\lambda_1)$. Hence, 
$[t(\lambda_1), \ t(\lambda_2)]=0$.

\section{ 
\rouge{ %% 
  Soliton non-preserving case
}}
\label{sect:SNP}

In this section, we consider the case where the reflection on the
boundary of the open chain also exchanges the fundamental
representation and its conjugate  (soliton non preserving case). 
The monodromy matrix itself is changed and includes alternating
fundamental-conjugate vector spaces along the chain (which is supposed
to have an even length $2L$).
\bleu{
\begin{eqnarray}
  T_{a}(\lambda) &=& R_{a\,2L}(\lambda) \bar R_{a\,2L-1}(\lambda) \ldots
  R_{a\,2}(\lambda) \bar R_{a\,1}(\lambda) 
\end{eqnarray}
}
$$\includegraphics{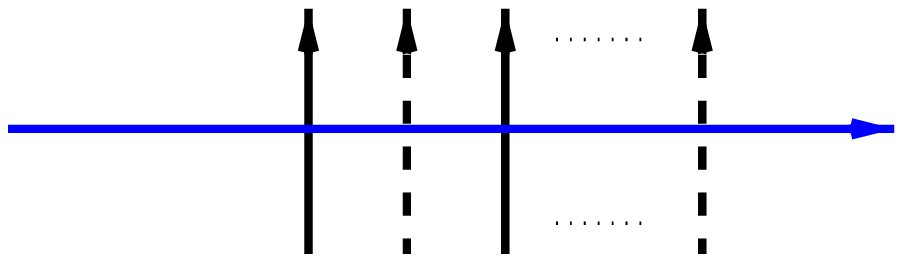}
$$
with \dgreen{$\bar R(\lambda) = R^{t_1}(-\lambda -i\rho) =
  R^{t_2}(-\lambda -i\rho)$} and $2\rho =
\theta_0(\emme -\enne)$, $\theta_0=\pm 1$. 
We use a transposition $^t$ which is related to the usual
transposition $^T$ by 
($A$ is any matrix):
\begin{equation}
A^t=V^{-1}\,A^T\,V \qmbox{where}\left\{
\begin{array}{ll}
  V = \mbox{antidiag}(1,1,\ldots,1)\,, &\ 
  \\ \qquad \mbox{ for which }\
  V^2=\theta_0=1\\
  \mbox{or}&\\
  V=\mbox{antidiag}\Big(\,
  \underbrace{1,\ldots,1}_{\enne/2}\,,\,\underbrace{-1,\ldots,-1}_{\enne/2}\,
    \Big)\,, \\
    \qquad \mbox{ for which }\ 
  V^2=\theta_0=-1\,.
\end{array}\right.
\label{eq:V}
\end{equation}
The second case is forbidden for $\enne$ odd.
\\
\bleu{
\begin{eqnarray}
  \hat T_{\bar a}(\lambda) &=& R_{1\,a}(\lambda)\bar R_{2\,a}(\lambda) \ldots
  R_{2L-1\,a}(\lambda) \bar R_{2L\,a}(\lambda) 
\end{eqnarray}
}
$$\includegraphics{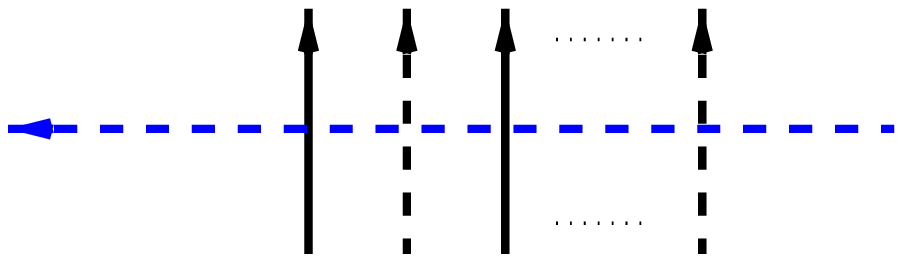}
$$
The two-line transfer matrix for the open chain with soliton
non-preserving boundary conditions is then defined by
\bleu{
\begin{equation}
  t(\lambda) = \tr_a \tK_{a}^{+}(\lambda)\ T_{a}(\lambda)\
  \tK_{a}^{-}(\lambda)\ \hat T_{\bar a}(\lambda) 
\end{equation} 
}
$$\includegraphics{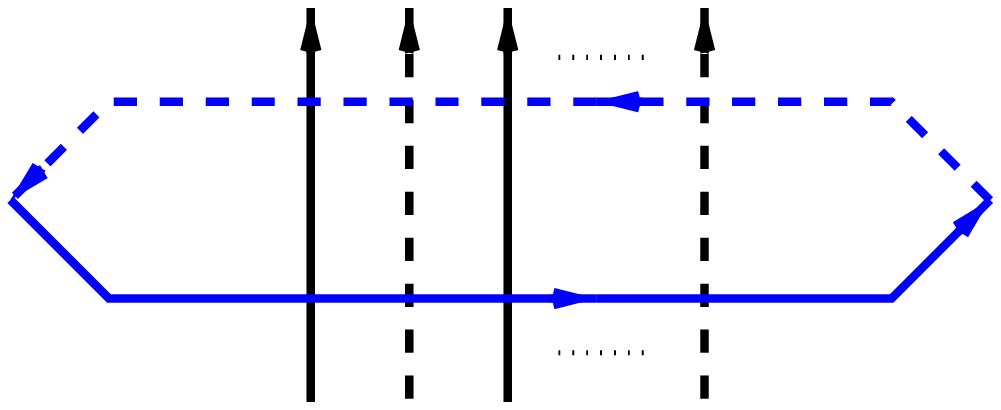}
$$
where $\tr_{a}$ denotes here the {\it super} trace
over the auxiliary space.

The commutation of transfer matrices for different values of the
spectral parameter now relies on the (local) reflection equation
\begin{equation}
  R_{ab}(\lambda_{a}-\lambda_{b})\ \tK_{a}(\lambda_{a})\ \bar
  R_{ba}(\lambda_{a}+\lambda_{b})\ \tK_{b}(\lambda_{b}) =
  \tK_{b}(\lambda_{b})\ \bar R_{ab}(\lambda_{a}+\lambda_{b})\
  \tK_{a}(\lambda_{a})\ R_{ba}(\lambda_{a}-\lambda_{b}).
\label{re2}
\end{equation}

\section{ 
\rouge{ %% 
  Solutions of the reflection equation
}}
\label{sect:RE}

\subsection{ 
\dcyan{ %% 
  Solutions to the soliton-preserving reflection equation
}}

Any bosonic invertible solution of the soliton preserving 
reflection equation (RE)
\newcommand{\EE}{{\mathbb E}}
\bleu{
\begin{eqnarray}
  && R_{12}(\lambda_{1}-\lambda_{2})K_{1}(\lambda_{1})
  R_{12}(\lambda_{1}+\lambda_{2}) K_{2}(\lambda_{2})=
  \nonumber\\[3mm]
  &&\qquad \qquad
  K_{2}(\lambda_{2})R_{12}(\lambda_{1}+\lambda_{2})
  K_{1}(\lambda_{1})R_{12}(\lambda_{1}-\lambda_{2})
\end{eqnarray}
}
where $R_{12}(\lambda)=\lambda\,\II+i\,P_{12}$ is the super-Yangian
$R$-matrix, is of the form 

\rouge{
\begin{equation}
  K(\lambda) = U\,\left( i\xi\,\II
    +\lambda\,\EE\right)U^{-1}
\end{equation}
}
where $U$ is independent of $\lambda$ and either
\begin{itemize}
\item[(i)]
  $\EE$ is diagonal and \rouge{$\EE^2=\II$} (diagonalisable solutions)
\item[(ii)]
  $\EE$ is strictly triangular and \rouge{$\EE^2=0$}
  (non-diagonalisable solutions)
\end{itemize}

\subsection{ 
\dcyan{ %% 
  Solutions to the soliton NON-preserving reflection equation
}}
Any bosonic invertible solution of the soliton non-preserving RE 
\bleu{
\begin{eqnarray}
  && R_{12}(\lambda_{1}-\lambda_{2})\ \tK_{1}(\lambda_{1})\ 
  R_{21}^{t_1}(\lambda_{1}+\lambda_{2})\ \tK_{2}(\lambda_{2}) = 
  \nonumber\\[3mm]
  &&\qquad \qquad
  \tK_{2}(\lambda_{2})\  R_{12}^{t_1}(\lambda_{1}+\lambda_{2})\
  \tK_{1}(\lambda_{1})\ R_{21}(\lambda_{1}-\lambda_{2}) 
  \label{re2-encore}
\end{eqnarray}
}
where $R_{12}(\lambda)=\lambda\,\II+i\,P_{12}$ is the super-Yangian
$R$-matrix, 
is a constant matrix  
(up to a multiplication by a scalar function) such that 
\rouge{
$\tK^t = \pm \tK$}.

\section{ 
\rouge{ %% 
   Pseudovacuum and one eigenvalue of the transfer matrix
}}
\label{sect:Lambda0}

We now choose an appropriate pseudo-vacuum, which is an exact
eigenstate of the transfer matrix :
\bleu{
\begin{eqnarray}
\vert \omega_{+} \rangle = \bigotimes_{i=1}^{L} \vert + \rangle
_{i} ~~~~\mbox{where} ~~~~\vert + \rangle = \left (
\begin{array}{c}
1 \\
0 \\
\vdots \\
0 \\
\end{array}
\right)\,\in\,\CC^{M+N}\,. \label{pseudo}
\end{eqnarray}
}
i.e.
{
\rouge{
  \begin{eqnarray}
    t(\lambda) ~\vert ~\omega_{+} \rangle  
    &=& \Lambda^0(\lambda) ~\vert \omega_{+} \rangle 
  \end{eqnarray}
}
}
with
\bleu{
\begin{eqnarray}
  \label{eq:eigen0}
  \Lambda^{0}(\lambda) &=& \alpha(\lambda)^{L} g_{0}(\lambda) +
  \beta(\lambda)^{L} 
  \sum_{l=1}^{{M}+{N}-2} (-1)^{[l+1]}
  g_{l}(\lambda)
  \\[3mm]&&
  + \gamma(\lambda)^{L} (-1)^{[{M}+{N}-1]}
  g_{{M}+{N}-1}(\lambda)
\end{eqnarray}
}
where, using
\begin{align}
  &a(\lambda) =\lambda+i \;, 
  &&b(\lambda)=\lambda \;, \\
  &\bar a(\lambda) =
  a(-\lambda - i\rho ) \;, \qquad
  &&\bar b(\lambda) = b(-\lambda -i\rho)  \;,
  \label{abaBbB}
\end{align} 
the functions $\alpha$, $\beta$, $\gamma$ and $g_l$ are defined as:\\[2mm]
\green{
\textit{(i) Soliton preserving boundary conditions with $L$ sites}
}
\dcyan{
\begin{eqnarray}
  && \alpha(\lambda) = a^{2}(\lambda),
  \qquad \beta(\lambda)=\gamma(\lambda)=b^{2}(\lambda)
\end{eqnarray}
}
and
\bleu{
\begin{eqnarray}
  \label{eq:glambda}
  && g_{l}(\lambda) = \frac{\lambda(\lambda + \frac{i({M}-{N})}{2})}
  {(\lambda + \frac{il}{2})(\lambda + \frac{i(l+1)}{2})} \;, \quad l =
  0,\ldots,{M}-1 \nonumber \\
  && g_{l}(\lambda) = \frac{\lambda(\lambda + \frac{i({M}-{N})}{2})}
  {(\lambda + \frac{i(2{M}-l-1)}{2})(\lambda + \frac{i(2{M}-l)}{2})}
  \;, \quad l \ge {M}
\end{eqnarray}
}
%$\rho = {{M} -{N} \over 2}$ in this convention. 
\\
\green{
\textit{(ii) Soliton non--preserving boundary conditions with $2L$ sites} 
}
\\
The basis used until now was the distinguished basis of $sl(M|N)$,
where the indices of $sl(M)$ come in first place, $1,\cdots,M$ and
those of $sl(N)$ afterwards $M+1,\cdots,M+N$. In the soliton non
preserving case we take $N=2n$ even. 
We consider in that case the symmetric basis for $sl(M|2n)$, where
the $2n$ indices of $sl(2n)$ are split in two parts: $1,\cdots,n$ and
$n+M+1,\cdots,n+M+n$, whereas the indices of the $sl(M)$ part are in
the middle: $n+1,\cdots,M+n$.
(see \cite{selene} for details)
\dcyan{
\begin{eqnarray}
  && \alpha(\lambda) = \Big(a(\lambda) \bar b(\lambda)\Big)^2,
  ~~\beta(\lambda) = \Big(b(\lambda)\bar b(\lambda)\Big)^2,
  \\[3mm]&&
   \gamma(\lambda) = \Big(\bar a(\lambda) b(\lambda)\Big)^2
\end{eqnarray}
}
and
\bleu{
\begin{eqnarray}
  g_{l}(\lambda) &=& {\lambda + {i\over 2}(\rho -1) \over \lambda +
  {i\rho \over 2}}, \qquad 0\leq l<{{M} +{N} -1 \over 2}\nonumber \\[3mm]
  g_{{M +{N}-1 \over 2}}(\lambda) &=&1,\qquad 
  \qmbox{if}M +{N} \qquad \mbox{odd}
  \nonumber \\ 
  g_{l}(\lambda) &=& g_{N+{M}-l-1}(-\lambda-i\rho). \qquad 
  \label{gb} 
\end{eqnarray}
}

\section{Analytical Bethe Ansatz}
\label{sect:ABA}

The other eigenvalues are supposed to be obtained by ``dressing'' with
rational functions
\bleu{
\begin{eqnarray}
  \Lambda(\lambda) &=& \alpha(\lambda)^{L} g_{0}(\lambda)
  A_0(\lambda) 
  +
  \beta(\lambda)^{L} 
  \sum_{l=1}^{{M}+{N}-2} (-1)^{[l+1]}
  g_{l}(\lambda) A_l(\lambda)
  \nonumber\\[3mm]&&
  + \gamma(\lambda)^{L} (-1)^{[{M}+{N}-1]}
  g_{{M}+{N}-1}(\lambda) A_{M+N-1}(\lambda)
\end{eqnarray}
}

\subsection{Bethe Ansatz equations in the soliton preserving case}

{From} the analyticity  of $\Lambda(\lambda)$,  one gets
\begin{eqnarray}
  \label{eq:anal1}
  A_{l}(-\frac{il}{2}) &=& A_{l-1}(-\frac{il}{2}), \qquad
  ~l=1,\ldots,{\emme}-1, \nonumber \\
  A_{2{\emme}-l}(-\frac{il}{2}) &=& A_{2{\emme}-l-1}(-\frac{il}{2}), \qquad
  l={\emme}-{\enne}+1,\ldots,{\emme}-1
\end{eqnarray}
Gathering together
all the constraints one can
determine the dressing functions, i.e.
\begin{eqnarray}
  A_{0}(\lambda) &=& \prod_{j=1}^{M^{(1)}} {\lambda+
  \lambda_{j}^{(1)}-\frac{i}{2}\over \lambda+ \lambda_{j}^{(1)} +\frac{i}{2}
  }\ {\lambda-\lambda_{j}^{(1)}-{i\over 2} \over \lambda-\lambda_{j}^{(1)}
  +{i\over 2}} \nonumber \\
  A_{l}(\lambda) &=& \prod_{j=1}^{M^{(l)}}
  {\lambda+\lambda_{j}^{(l)}+{il\over 2}+i \over \lambda+ \lambda_{j}^{(l)}
  +{il\over2}} \; {\lambda-\lambda_{j}^{(l)}+{il\over 2}+i\over \lambda-
  \lambda_{j}^{(l)} +{il\over 2}} \nonumber \\
  && \times \prod_{j=1}^{M^{(l+1)}}{\lambda+ \lambda_{j}^{(l+1)}+{il\over
  2}-{i\over 2}\over \lambda+ \lambda_{j}^{(l+1)} +{il\over 2} +{i\over 2}}\
  {\lambda-\lambda_{j}^{(l+1)}+{il \over 2}-{i\over 2} \over
  \lambda-\lambda_{j}^{(l+1)} + {il\over 2}+{i\over 2}} \qquad l =
  1,\ldots,{\emme}-1 \nonumber \\
  A_{l}(\lambda) &=& \prod_{j=1}^{M^{(l)}}
  {\lambda+\lambda_{j}^{(l)}+i{\emme}-{il\over 2}-i \over \lambda+
  \lambda_{j}^{(l)} +i{\emme}-{il\over2}} \;
  {\lambda-\lambda_{j}^{(l)}+i{\emme}-{il\over 2}-i\over \lambda-
  \lambda_{j}^{(l)} +i{\emme}-{il\over 2}} \nonumber \\
  && \times \prod_{j=1}^{M^{(l+1)}}{\lambda+
  \lambda_{j}^{(l+1)}+i{\emme}-{il\over 2}+{i\over 2}\over \lambda+
  \lambda_{j}^{(l+1)} +i{\emme}-{il\over 2} -{i\over 2}}\
  {\lambda-\lambda_{j}^{(l+1)}+i{\emme}-{il \over 2}+{i\over 2} \over
  \lambda-\lambda_{j}^{(l+1)} +i{\emme}- {il\over 2}-{i\over 2}} \nonumber
  \\
  &&l = {\emme},\ldots,{\emme}+{\enne}-1
  \label{eq:dressingslmn}
\end{eqnarray}
Analyticity around the poles introduced in the factors $A_l$ finally
imposes the so-called Bethe equations in the $\lambda_i$ :
\begin{eqnarray}
  e_{1}(\lambda_{i}^{(1)})^{2L} &\!\!=\!\!& -\prod_{j=1}^{M^{(1)}}
  e_{2}(\lambda_{i}^{(1)} - \lambda_{j}^{(1)})\ e_{2}(\lambda_{i}^{(1)} +
  \lambda_{j}^{(1)})\ 
  \nonumber\\ 
  && \times \prod_{ j=1}^{M^{(2)}}e_{-1}(\lambda_{i}^{(1)} -
  \lambda_{j}^{(2)})\ e_{-1}(\lambda_{i}^{(1)} + \lambda_{j}^{(2)})\,,
  \nonumber \\
  1 &\!\!=\!\!& -\prod_{j=1}^{M^{(l)}} e_{2}(\lambda_{i}^{(l)} -
  \lambda_{j}^{(l)})\ e_{2}(\lambda_{i}^{(l)} + \lambda_{j}^{(l)})\ 
  \nonumber\\ 
  && \times \prod_{
  \tau = \pm 1}\prod_{ j=1}^{M^{(l+\tau)}}e_{-1}(\lambda_{i}^{(l)} -
  \lambda_{j}^{(l+\tau)})\ e_{-1}(\lambda_{i}^{(l)} +
  \lambda_{j}^{(l+\tau)}) 
  \nonumber \\
  && l= 2,\ldots,{\emme}-1,{\emme}+1,\ldots,{\emme}+{\enne}-2 
  \nonumber \\
  1 &\!\!=\!\!& 
  \prod_{ j=1}^{M^{({\emme}-1)}}e_{-1}(\lambda_{i}^{({\emme})} -
  \lambda_{j}^{({\emme}-1)})\ e_{-1}(\lambda_{i}^{({\emme})} +
  \lambda_{j}^{({\emme}-1)}) 
  \nonumber \\
  &&\times 
  \prod_{j=1}^{M^{({\emme}+1)}} e_{1}(\lambda_{i}^{({\emme})}
  - \lambda_{j}^{({\emme}+1)})\ e_{1}(\lambda_{i}^{({\emme})} +
  \lambda_{j}^{({\emme}+1)})\ 
  \nonumber 
\end{eqnarray}
\begin{eqnarray}
  1 &\!\!=\!\!& -
  \prod_{j=1}^{M^{({\emme}+{\enne}-2)}}e_{-1}(\lambda_{i}^{({\emme}+{\enne}-1)}
  - \lambda_{j}^{({\emme}+{\enne}-2)})\
  \nonumber\\ && \qquad\qquad\quad
   e_{-1}(\lambda_{i}^{({\emme}+{\enne}-1)} +
  \lambda_{j}^{({\emme}+{\enne}-2)}) \nonumber \\
  &&\times \prod_{j=1}^{M^{({\emme}+{\enne}-1)}}
  e_{2}(\lambda_{i}^{({\emme}+{\enne}-1)} -
  \lambda_{j}^{({\emme}+{\enne}-1)})\
  \nonumber\\ && \qquad\qquad\quad
  e_{2}(\lambda_{i}^{({\emme}+{\enne}-1)} +
  \lambda_{j}^{({\emme}+{\enne}-1)})
  \label{BAE}
\end{eqnarray}
with 
\dgreen{
\begin{equation}
  e_{x}(\lambda)=\frac{\lambda+\frac{ix}{2}}{\lambda-\frac{ix}{2}}
\end{equation}
}

We now 
implement non trivial soliton preserving boundary conditions $K^-$. From
the classification given in section \ref{sect:RE}, we know that
$K^-(\lambda)$
is always conjugated (by a constant matrix $U$) to a diagonal matrix of
the form
\begin{eqnarray}
  K(\lambda) = \diag( \underbrace{\alpha, \ldots ,\alpha}_{m_{1}},
  \underbrace{\beta, \dots, \beta}_{m_{2}}, \underbrace{\beta, \dots,
  \beta}_{n_{2}}, \underbrace{\alpha, \ldots ,\alpha}_{n_{1}} )
  \label{eq:solDiag}
\end{eqnarray}
Then, it is easy to see that the spectrum and the symmetry of
the model depend only on the diagonal, and not on
$U$. Indeed, when considering two reflection matrices related by a
constant conjugation, the corresponding transfer matrices are also
conjugated. Thus,  it is enough to
consider
diagonal $K^-(\lambda)$ matrices to get the general case. Such a
property, which relies on the form of the $R$-matrix, is a priori 
valid only in the \emph{rational} 
$sl(\enne)$ and $sl(\emme|{\enne})$ cases.
\\
For a diagonal solution 
with $m_{1}+m_{2}={\emme}$, $n_{1}+n_{2}={\enne}$, $\alpha(\lambda) =
-\lambda +i\xi$, $\beta(\lambda) = \lambda+i\xi$, and the free
boundary parameter $\xi$, one can compute the new form $\widetilde
g_{l}(\lambda)$  of the $g$-functions entering the expression of
$\widetilde\Lambda_{0}(\lambda)$, the new pseudo-vacuum eigenvalue.
They take the form:
\begin{eqnarray}
  \widetilde g_{l}(\lambda) &=& (-\lambda + i\xi)\, g_{l}(\lambda), \qquad
  l=0, \ldots, m_{1}-1 \nonumber \\
  \widetilde g_{l}(\lambda) &=& (\lambda + i\xi +im_{1})\, g_{l}(\lambda),
  \qquad l=m_{1}, \ldots, {\emme}+n_{2}-1 \nonumber \\
  \widetilde g_{l}(\lambda) &=& (-\lambda + i\xi - im_{2} + in_{2})\,
  g_{l}(\lambda), \qquad l={\emme}+n_{2}, \ldots, {\emme}+{\enne}-1
  \label{eq:tg1}
\end{eqnarray}
where $g_l(\lambda)$ are given by (\ref{eq:glambda}).
The dressing functions $A_l$ keep the same form, but
the Bethe Ansatz 
equations  are modified (by $K^{-}(\lambda)$), so that the
value of the 
eigenvalues $\Lambda(\lambda)$ are different from the ones obtained
when $K(\lambda)=\II$.

\medskip

The modifications induced on Bethe Ansatz equations 
are the following: \\
-- The factor $-e_{2\xi+m_{1}}^{-1}(\lambda)$ appears in the LHS of the
${m_{1}}^{th}$ Bethe equation. \\
-- The factor $-e_{2\xi+m_{1}-m_{2}-n_{2}}^{-1}(\lambda)$ appears in the
LHS of the $({\emme}+n_{2})^{th}$ Bethe equation.

\subsection{Bethe Ansatz equations in the soliton non preserving case}

The dressing functions now take the form:
\begin{eqnarray}
  A_{0}(\lambda) &=& \prod_{j=1}^{M^{(1)}}{\lambda+
  \lambda_{j}^{(1)}-{i\over 2}\over \lambda+ \lambda_{j}^{(1)} +{i\over 2}}\
  {\lambda-\lambda_{j}^{(1)}-{i\over 2} \over \lambda-\lambda_{j}^{(1)}
  +{i\over 2}} \,, \nonumber\\
  A_{l}(\lambda) &=& \prod_{j=1}^{M^{(l)}}
  {\lambda+\lambda_{j}^{(l)}+{il\over 2}+i \over \lambda+ \lambda_{j}^{(l)}
  +{il\over2}} \; {\lambda-\lambda_{j}^{(l)}+{il\over 2}+i\over \lambda-
  \lambda_{j}^{(l)} +{il\over 2}} \nonumber \\
  && \times \prod_{j=1}^{M^{(l+1)}}{\lambda+ \lambda_{j}^{(l+1)}+{il\over
  2}-{i\over 2}\over \lambda+ \lambda_{j}^{(l+1)} +{il\over 2} +{i\over 2}}\
  {\lambda-\lambda_{j}^{(l+1)}+{il \over 2}-{i\over 2} \over
  \lambda-\lambda_{j}^{(l+1)} + {il\over 2}+{i\over 2}} \,, \qquad l =
  1,\ldots , n-1 \label{dressingSNPslmn}\nonumber \\
  A_{l}(\lambda) &=& \prod_{j=1}^{M^{(l)}}
  {\lambda+\lambda_{j}^{(l)}+in-{il\over 2}-i \over \lambda+
  \lambda_{j}^{(l)} +in-{il\over2}} \;
  {\lambda-\lambda_{j}^{(l)}+in-{il\over 2}-i\over \lambda-
  \lambda_{j}^{(l)} +in-{il\over 2}} \nonumber \\
  && \times \prod_{j=1}^{M^{(l+1)}}{\lambda+
  \lambda_{j}^{(l+1)}+in-{il\over 2}+{i\over 2}\over \lambda+
  \lambda_{j}^{(l+1)} +in-{il\over 2} -{i\over 2}}\
  {\lambda-\lambda_{j}^{(l+1)}+in-{il \over 2}+{i\over 2} \over
    \lambda-\lambda_{j}^{(l+1)} +in- {il\over 2}-{i\over 2}} \,, 
  \\&&\qquad
  n\le l < n+\frac{\emme-1}2 \nonumber
\end{eqnarray}
and $A_{l}(\lambda) = A_{\emme+2n-1-l}(-\lambda -i\rho)$,
and for $\emme=2m+1$
\begin{eqnarray}
  A_{k}(\lambda) &=& \prod_{j=1}^{M^{(k)}}
  {\lambda+\lambda_{j}^{(k)}+in-{ik\over 2}-i\over \lambda+
  \lambda_{j}^{(k)} +in-{ik\over2}} \;
  {\lambda-\lambda_{j}^{(k)}+in-{ik\over 2}-i\over \lambda-
  \lambda_{j}^{(k)} +in-{ik\over 2}} \nonumber \\ 
  && \times{\lambda+\lambda_{j}^{(k)}+in-{ik\over 2}+{i\over 2} \over
  \lambda+ \lambda_{j}^{(k)} +in-{ik\over 2}-{i\over 2}} \;
  {\lambda-\lambda_{j}^{(k)}+in-{ik\over 2}+{i\over 2}\over \lambda-
    \lambda_{j}^{(k)} +in-{ik\over 2}-{i\over 2}}\,, 
  \\&& (k=m+n)\nonumber 
  \label{a21}
\end{eqnarray}
and the Bethe Ansatz equations read as:
 
\subsubsection*{A. $\bf sl(2m+1|2n)$ superalgebra} 

\begin{eqnarray}
  e_{1}(\lambda_{i}^{(1)})^{2L} &\!\!=\!\!& -\prod_{j=1}^{M^{(1)}}
  e_{2}(\lambda_{i}^{(1)} - \lambda_{j}^{(1)})\ e_{2}(\lambda_{i}^{(1)} +
  \lambda_{j}^{(1)})\ 
  \nonumber \\
  &&\times 
\prod_{ j=1}^{M^{(2)}}e_{-1}(\lambda_{i}^{(1)} -
  \lambda_{j}^{(2)})\ e_{-1}(\lambda_{i}^{(1)} + \lambda_{j}^{(2)})\,,
  \nonumber \\
  1 &\!\!=\!\!&- \prod_{j=1}^{M^{(l)}} e_{2}(\lambda_{i}^{(l)} -
  \lambda_{j}^{(l)})\ e_{2}(\lambda_{i}^{(l)} + \lambda_{j}^{(l)})\ 
  \nonumber \\
  &&\times 
\prod_{
  \tau = \pm 1}\prod_{ j=1}^{M^{(l+\tau)}}e_{-1}(\lambda_{i}^{(l)} -
  \lambda_{j}^{(l+\tau)})\ e_{-1}(\lambda_{i}^{(l)} +
  \lambda_{j}^{(l+\tau)}) \nonumber \\
  && l= 2,\ldots,n+m-1, \;\; l \neq n \nonumber \\
  1 &\!\!=\!\!& \prod_{j=1}^{M^{(n+1)}} e_{1}(\lambda_{i}^{(n)} -
  \lambda_{j}^{(n+1)})\ e_{1}(\lambda_{i}^{(n)} + \lambda_{j}^{(n+1)})\
  \nonumber \\
  &&\times 
  \prod_{ j=1}^{M^{(n-1)}}e_{-1}(\lambda_{i}^{(n)} - \lambda_{j}^{(n-1)})\
  e_{-1}(\lambda_{i}^{(n)} + \lambda_{j}^{(n-1)}) \nonumber \\
  e_{-{1\over 2}}(\lambda_{i}^{(k)}) &\!\!=\!\!& -\prod_{j=1}^{M^{(k)}}
  e_{2}(\lambda_{i}^{(k)} - \lambda_{j}^{(k)})\
  e_{2}(\lambda_{i}^{(k)} + \lambda_{j}^{(k)})\ 
  \nonumber \\ && \qquad\qquad
  e_{-1}(\lambda_{i}^{(k)} - \lambda_{j}^{(k)})\
  e_{-1}(\lambda_{i}^{(k)} + \lambda_{j}^{(k)})\ \nonumber \\
  &&\!\! \times\prod_{j=1}^{M^{(k-1)}}e_{-1}(\lambda_{i}^{(k)} -
  \lambda_{j}^{(k-1)})\ e_{-1}(\lambda_{i}^{(k)} +
  \lambda_{j}^{(k-1)}) \,, ~~k=m+n 
  \label{BAE1n}
\end{eqnarray}

Note that these equations are the
Bethe Ansatz equations of the $osp(2m+1|2n)$ case (see e.g. \cite{yabon}) 
apart from the last equation.
 
\subsubsection*{B. $\bf sl(2m|2n)$ superalgebra} 
 
The first $n+m-1$ equations are the same as in the previous case,  
but the last equation is modified, with again  $k=m+n$,  to
\begin{eqnarray}
  e_{1}(\lambda_{i}^{(k)}) &\!\!=\!\!& -\prod_{j=1}^{M^{(k)}}
  e_{2}(\lambda_{i}^{(k)} - \lambda_{j}^{(k)})\
  e_{2}(\lambda_{i}^{(k)} + \lambda_{j}^{(k)})\ 
  \nonumber \\
  &&\times 
  \prod_{j=1}^{M^{(k-1)}}e_{-1}^2(\lambda_{i}^{(k)} -
  \lambda_{j}^{(k-1)})\ e_{-1}^2(\lambda_{i}^{(k)} +
  \lambda_{j}^{(k-1)}) \,. \qquad
  \label{BAE2n}
\end{eqnarray}

\subsubsection*{\bf With a non trivial diagonal reflection matrix $K^-$}
In the case of a non trivial diagonal reflection matrix $K^-$
with $\eps=1$, i.e.
\begin{equation}
    \tK^{-}(\lambda)=\diag(k_{1},\ldots,k_{\emme+\enne})\qmbox{with}
    k_{\emme+\enne+1-j}=\,k_{j} \,,
\end{equation}
the $g$-functions entering the new pseudo-vacuum eigenvalue are
modified as:
\begin{eqnarray}
  \widetilde g_{l}(\lambda) &=& k_{l+1}\, g_{l}(\lambda), \qquad
  0\leq l \leq \frac{\emme+\enne-1}2 
\end{eqnarray}
where $g_l(\lambda)$ are given by (\ref{gb}).
The remaining $\widetilde g$ are defined by requiring the
crossing relation
\begin{equation}
   \widetilde  g_{\emme+\enne-l}(-\lambda-i\rho)= \, 
   \widetilde  g_{l}(\lambda)\,.
\end{equation}
The dressing functions (\ref{dressingSNPslmn}) and (\ref{a21})
keep the same form, but the LHS of $\ell^{th}$ Bethe Ansatz
equation (given in (\ref{BAE1n}) and (\ref{BAE2n})) is multiplied by
$k_{\ell}/k_{\ell+1}$.

\section*{Acknowledgments}
This work has been financially supported by the TMR Network
EUCLID: ``Integrable models and applications: from strings to
condensed matter'', contract number HPRN-CT-2002-00325.

\end{document}